\documentclass[12pt]{article}
\usepackage{graphicx}
\begin{document}

\centerline{\bf \large Sociophysics Simulations II: Opinion Dynamics}

\bigskip
\centerline{Dietrich Stauffer}

\bigskip
\centerline{Institute for Theoretical Physics, Cologne University, D-50923 K\"oln, Euroland}

\begin{abstract}
Individuals have opinions but can change them under the influence of others.
The recent models of Sznajd (missionaries), of Deffuant et al. (negotiators), 
and of Krause and Hegselmann (opportunists)
are reviewed here, while the voter and Ising models, Galam's majority rule and 
the Axelrod multicultural model were dealt with by other lecturers at this 
8th Granada Seminar. 
\end{abstract}

\section{Introduction}
University professors know everything and are always right; lesser people 
change their opinion after interactions with others, as discussed in this 
review as well as in the lectures of Redner, Toral, and San Miguel. Missing at 
this Granada seminar was the Latan\'e model \cite{nowak,kohring} which is a
generalized Ising model; simulations are reviewed in \cite{holyst}. Here 
we concentrate on the models of Sznajd \cite{sznajd}, Krause and Hegselmann
\cite{krause}, and Deffuant et al \cite{deffuant}, all three of which differ 
drastically in their definitions but give similar results, just as many variants
of the Sznajd model were shown to have similar properties in an analytical 
approximation \cite{galam}. For completeness we mention that the language
bit-string models of part I of our review series can also be interpreted as 
binary Axelrod models for multi-culturality: instead of taking over elements of
another language, one may also replace elements of  the native culture by those
of another culture. On the other hand, the Latan\'e model was already applied 
to languages in \cite{nettle}.
 
In the next section we first define the three models in a unified way, and then 
present, section by section, selected results. 

\begin{figure}[hbt]
\begin{center}
\includegraphics[angle=-90,scale=0.5]{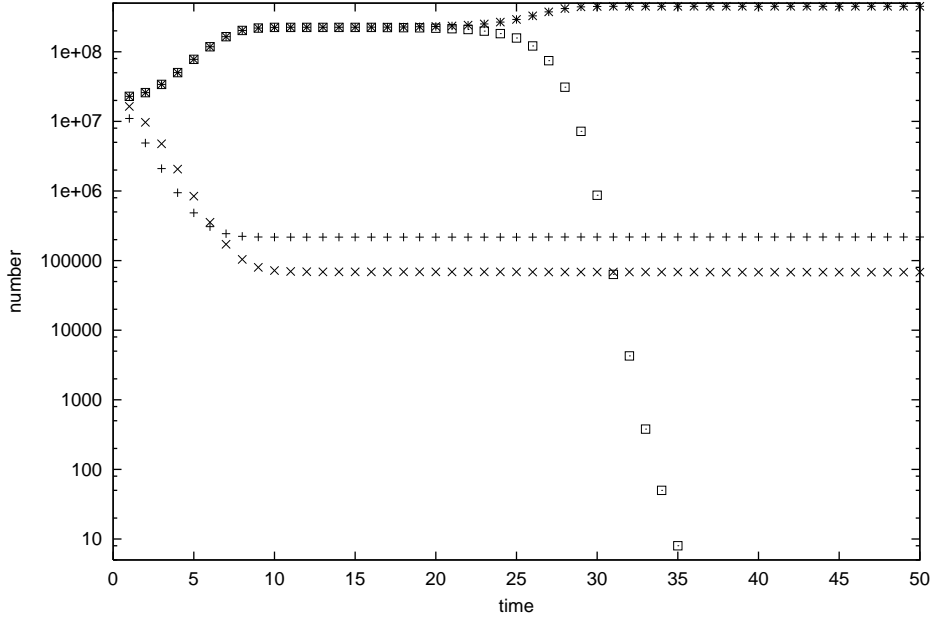}
\end{center}
\caption{Continuous opinions for 450 million negotiators with everybody 
possibly connected to everybody: Two centrist parties fight for victory while 
much smaller extremist parties survive unharmed. From \cite{forsta}.
}
\end{figure}

\begin{figure}[hbt]
\begin{center}
\includegraphics[angle=-90,scale=0.5]{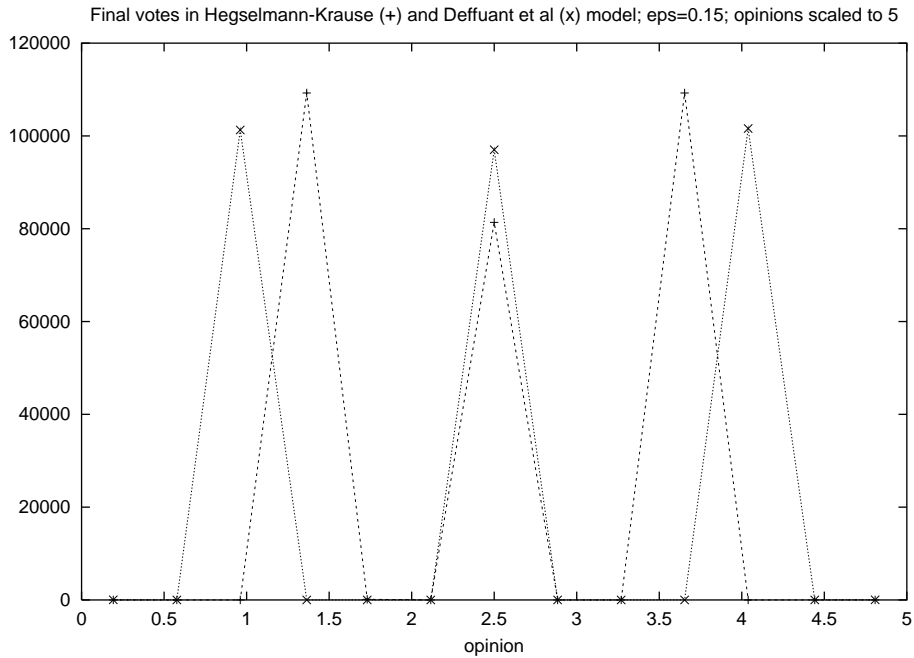}
\end{center}
\caption{Final distribution of votes for 300,000 opportunists (dashed line)
and negotiators (solid line), with opinions scaled from 0 to 5 instead
of the original interval [0,1] to facilitate comparison with missionaries, 
Fig.3. 
}
\end{figure}
  
\begin{figure}[hbt]
\begin{center}
\includegraphics[angle=-90,scale=0.5]{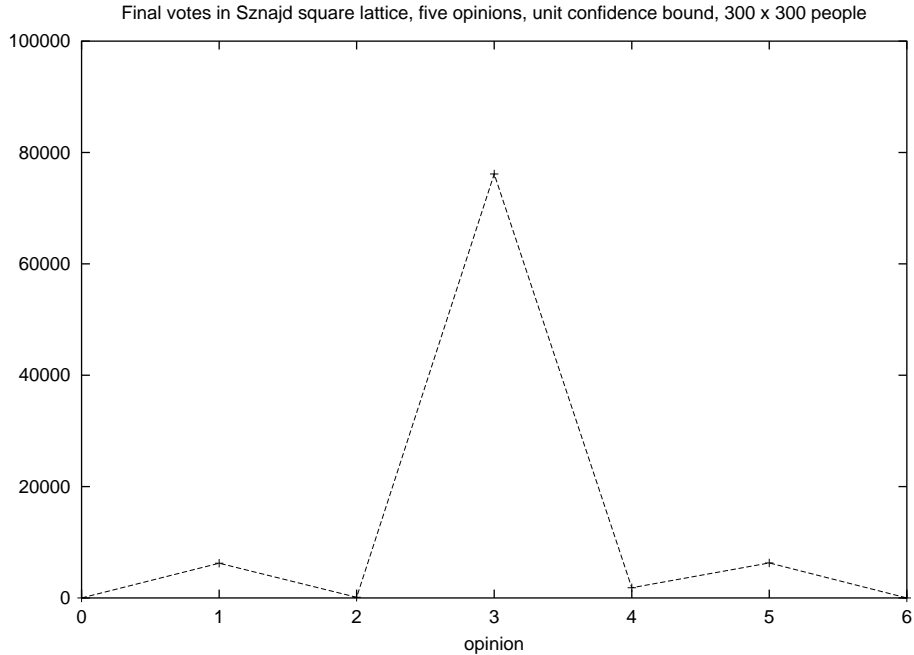}
\end{center}
\caption{Final distribution of votes for $301 \times 301$ missionaries
with discrete opinions 1, 2, 3, 4, 5 and unit confidence bound. Thus two 
neighbours of opinion 4 convince all those neighbours to the same opinion 4 
which had opinions 3 or 5 before. 
}
\end{figure}
  
\begin{figure}[hbt]
\begin{center}
\includegraphics[angle=-90,scale=0.5]{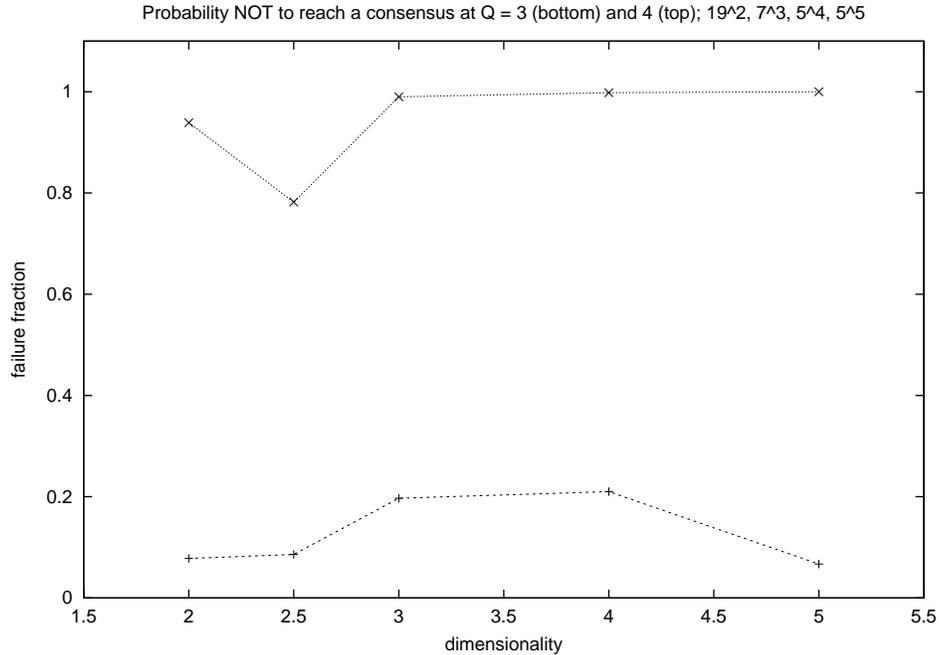}
\end{center}
\caption{Probability to find no consensus with three (lower data) and four
(upper data) possible missionary opinions, versus dimensionality. The triangular
lattices is put at dimensionality 5/2. Opinion $0$ can only convince opinions
$) \pm 1$; all opinions are natural numbers. From \cite{forsta}.  
}
\end{figure}

\begin{figure}[hbt]
\begin{center}
\includegraphics[angle=-90,scale=0.5]{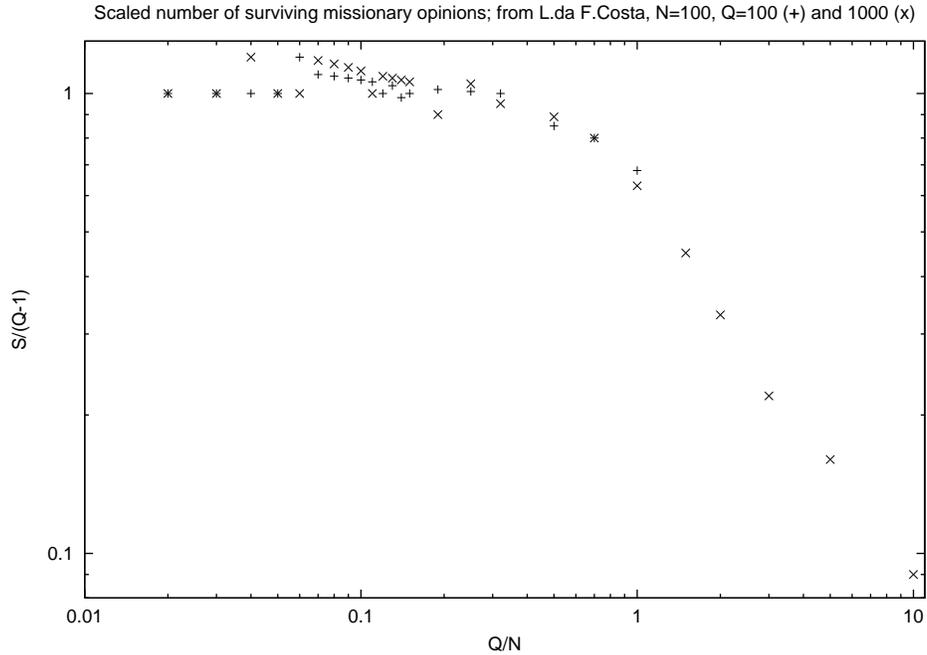}
\end{center}
\caption{Scaling of the number $S$ of surviving opinions as a function of the 
number $Q$ of possible discrete opinions and the number $N$ of people, for
missionaries. From L. da Fontura Costa, priv, comm. PRELIMINARY
}
\end{figure}

\begin{figure}[hbt]
\begin{center}
\includegraphics[angle=-90,scale=0.5]{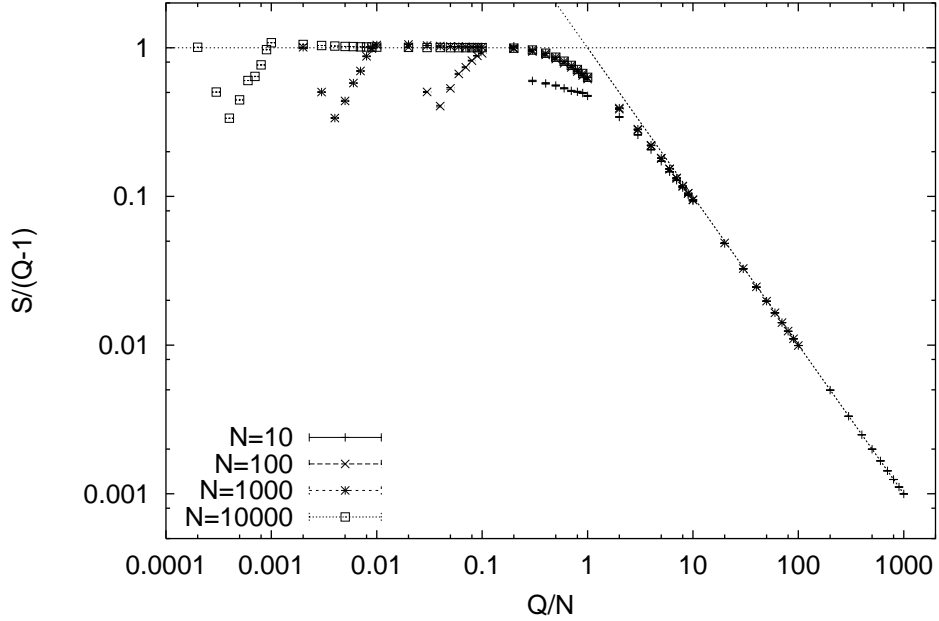}
\end{center}
\caption{As Fig.5 but for opportunists. From \cite{fort3}.
}
\end{figure}

\begin{figure}[hbt]
\begin{center}
\includegraphics[angle=-90,scale=0.5]{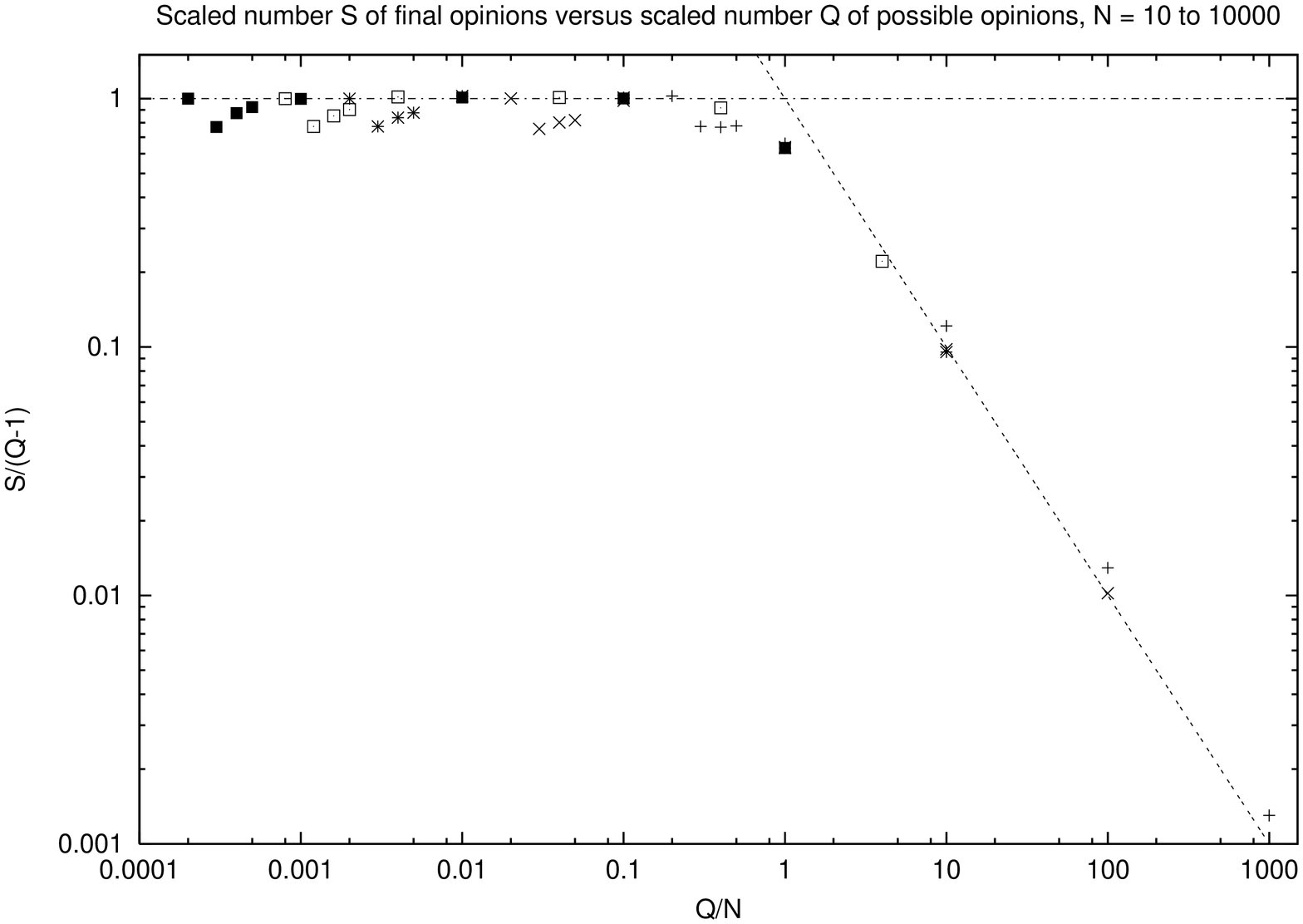}
\end{center}
\caption{As Fig.5 but for negotiators, from \cite{deffdis}. 
}
\end{figure}

\section {The Three Models}

Each individual $i\; (i=1,2, \dots N)$ has one opinion $O_i$ on one 
particular question. This opinion can be binary (0 or 1), multivalued integer
($O_i = 1, 2, ..., Q$) or continuous real ($0 \le O_i \le 1$). The neighbours $j$
of individual $i$ may be those on a square lattice, or on a Barab\'asi-Albert 
network, or any other individual. Because of interactions between individuals 
$i$ and $j$, one of them or both may change opinion from one time step ($t$) 
to the next ($t+1$), according to rules to be specified below.

``Bounded confidence'' \cite{krause,deffuant} means that only people with 
similar opinions talk to each other. If in politics five parties 1, 2, 3, 4, 5 
sit in parliament, traditionally ordered from left to right, then a left-centre
coalition of 2 and 3, or a rightist coalition of 3, 4, and 5 may work, while
collaboration of the extremes 1 and 5 seldomly happen in formal coalition 
agreements. Thus we may assume that only parties talk to each other which differ
by not more than one opinion unit, or by $\epsilon Q$ units more generally 
for $Q$ opinions, or by $\epsilon$ for real opinions between zero and one. If
$\epsilon \ge 1$, bounded confidence is ignored; if $\epsilon \ll 1$, confidence
is strongly bounded. This parameter $\epsilon$ thus measures the tolerance for
dissent or the openness to different opinions. 

\bigskip
The three models (each of which was studied in several variants) are 
missionaries \cite{sznajd}, opportunists \cite{krause} and negotiators 
\cite{deffuant}. 

{\bf Missionaries} of the Sznajd model convince all neighbours (within their 
confidence bound) of their mission, particularly if two neighbouring 
missionaries have the same opinion. For example, if on a square lattice two
neighbours have the same opinion $O = 2$ out of $Q = 5$ possible opinions, and
the confidence bound is one unit, then they force their opinion 2 onto all 
(at most six) lattice neighbours which before had opinions 1, 2, or 3; they 
cannot convince neighbours with opinions 4 or 5. 

{\bf Opportunists} of the Krause-Hegselmann model ask all their neighbours 
(within their confidence bound) for their opinion, and then follow the 
arithmetic average of them. Thus for $Q = 5$, a present opinion $O_i(t) = 2$ 
of the considered individual and a confidence bound of one unit, the new 
$O_i(t+1)$ will be the rounded arithmetic average of all neighbour opinions 
except 4 and 5. 

{\bf Negotiators} of Deffuant et al. each select one discussion partner at one 
time step. If their opinions $O_i$ and $O_j$ differ by less than the confidence
bound, their two opinions mutually get closer without necessarily agreeing 
completely. More precisely, $O_i$ shifts towards $O_j$ and $O_j$ shifts towards
$O_i$ by a (rounded) amount $\mu |O_j-O_i|$, where the extreme case $\mu = 0$ 
means rigid unchanging opinions, while $\mu = 1/2$ gives immediate agreement.
For example, for $Q=5$ and $\mu = 0.3$, for a confidence bound of three units, 
the pair $O_i=2, \; O_j=5$ will become $O_i=3, \; O_j=4$. (If the opinion 
difference is only one unit, one of the two partner takes the opinion of the 
other. Thus a confidence bound of only one unit makes less sense since then 
only this special case of one opinion jumping to the other remains, and 
no mutual compromise as in negotiations.) Already two centuries ago,
the mathematician Gauss (according to U. Krause) studied a similar problem: 
How do two opinions evolve if one discussion partner takes the arithmetic
and the other takes the geometric average of the two opinions. 

Particularly large populations can be simulated for the continuous negotiator
model of infinite connection range, Fig.1. To plot the continuous opinions
we binned them into 20 intervals and show only the centrist intervals 10 and 11
and the extremists in intervals 1 (+) and 2 (x). More plots on the time
dependence of negotiations are given in \cite{deffuant}.

Basic programs for missionaries, opportunists and negotiators are published in
my earlier reviews \cite{stauffer}.

\section{Consensus, Polarization or Fragmentation}

All three standard models give after sufficiently long time one of three types 
of results: We may find one, two or more than two different opinions surviving.
The case of one opinion or consensus can also be called dictatorship. The case
of two surviving opinions or polarization can also be called a balance of power 
between opposition and government. The case with three or more opinions 
or fragmentation can also be called anarchy, multi-party democracy, 
multiculturality or diversity. Thus the models themselves do not tell us whether
the result is desirable or unwanted; this value judgement depends on the 
application and interpretation. 

(Similarly, once we physicists have mastered
the multiplication $3 \times 5 = 15$ we can estimate that three bags, of five 
oranges each, contain in total 15 oranges, or that a room of 5 meter
length and 3 meter width has an area of 15 square meters. Both results are 
usually regarded as correct even though an orange does not have a surface of a 
square meter and a room of 15 square meters may be regarded as too large for
some and too small for other purposes. Thus one model, here multiplication,
may have different interpretations which can be judged differently.) 

Figs.2 and 3 show this similarity: Continuous opinions for opportunists and for 
negotiators give three final main opinions in Fig.2, and discrete opinions for
missionaries do the same in Fig.3. The distribution of people among the three
opinions may be different, with very tiny groups having opinions between the
main ones, or fringe opinions near zero and one not dying out for the 
continuous case. 

For missionaries, not only square lattices have been simulated. The original
one-dimensional chains are less interesting (similar to Ising models) since 
they do not have a phase transition (see next section). But for triangular,
simple cubic, and hypercubic in four and five dimensions the results are about
the same, Fig.4: For $Q \le 3$ possible opinions, in most cases a consensus 
is reached; for $Q \ge 4$ possible opinions, such a consensus is rare.
(The confidence bound is unity for all cases.) 
 
This threshold of 3.5 at unit confidence interval (or $\epsilon = 1/3.5$) for 
missionaries corresponds to a threshold of $\epsilon = 1/2$ for negotiators 
\cite{fort1} and $\epsilon = 0.2$ for opportunists \cite{fort2} with continuous
opinions between zero and one: For larger 
$\epsilon$ one has consensus, for decreasing epsilon one has first polarization 
into two opinions, and then fragmentation into three or more opinions,         
$\propto 1/\epsilon$. The negotiator threshold
1/2 is quite general \cite{fort1} except if the model is made very asymmetric
\cite{assmann}. In summary: Reaching a consensus requires a strong willingness
to listen to other opinions and to reach a compromise. For negotiators the
results are well described by a theory \cite{bennaim}.

Missionaries with continuous opinions seem to reach always a consensus 
\cite{fort4}, independent of the confidence bound $\epsilon$.

Discrete instead of continuous opinions have the advantage that one can find
precisely whether or not two opinions agree, without a numerical cutoff
depending on the precision of the computer. Particularly opportunists could now
be simulated in much larger numbers \cite{fort3}.  Also, now a fixed point is
reached when all opinions agree or are out of reach from each other; real 
numbers never fully agree and thus prevent a fixed point. Moreover, now one has 
a maximum number $Q$ of possible opinions, and the following scaling law is 
valid: If the number $N$ of people is much larger than the number $Q$ of 
possible 
opinions, then each opinion will find some followers, and the number $S$ of 
surviving opinions agrees with $Q$. In the opposite limit of $N \ll Q$, each
person may keep its own opinion if separated by more then $\epsilon$ from the
other opinions: $S = N$. It is easiest to take a unit confidence interval, 
i.e. $\epsilon = 1/Q$. Then for missionaries (Fig.5), opportunists (Fig.6) 
and negotiators (Fig.7) we get
$$ S/Q = f(Q/N) $$
with a constant scaling function $f$ for $Q \ll N$, and $f = N/Q$ for $Q \gg N$,
valid for large $Q, S, N$. 

\section{Networks}

Most simulations of opportunists and negotiators had infinite connection range, 
i.e.  each person could get into contact with all other persons, with the same 
probability. In contrast, the missionaries were usually simulated on lattices.
Reality is in between these two extremes of nearest lattice neighbours and
infinitely distant neighbours. Small world networks \cite{elgazzar} and 
in particular scale-free networks of Barab\'asi-Albert type \cite{ba} have been
used as the topological basis of opinion dynamics. The name scale-free means
that there is no characteristic number $k$ of neighbours for each site; instead
the number of sites having $k$ neighbours decays with a power-law in $k$ like
$1/k^3$. These networks are supposed to describe the empirical fact than with
a rather small number of steps one can connect most people in the USA with
most other people there via personal acquaintances.

Their history, often misrepresented,
started in 1941 with Flory's percolation theory on a Bethe 
lattice where each site has exactly $k$ neighbours, with the same $k$ for 
all sites; if the probability for two neighbours to be connected is larger than
the percolation threshold $p_c = 1/(k-1)$, one infinite cluster of connected 
sites appears, coexisting with many small clusters including isolated sites.
15 years later Erd\"os and R\'enyi modified it such that each site is connected
with a small probability with other sites, arbitrarily far away; this random 
graph belongs to the same ``universality class'' of mean-field percolation as 
Flory's solution but now the number $k$ of neighbours for each site fluctuates 
according to a Poisson distribution. We get the desired $1/k^3$ law only by a 
very
different construction (the rich get richer; or powerful people attract even 
more supporters). We start with $m$ sites all connected with each other. This
network is then enlarged, adding one site per step. Each new site selects,
among the already existing sites, exactly $m$ neighbours, randomly with a 
probability proportional to the number of neighbours that site had already 
before. Once a new site has selected an old site as a neighbour, also the new
site is neighbour to the old site: this neighbourhood relation is symmetric. 
A computer program published in \cite{deffdis} at the beginning contains this
construction of the network. As a result, the probability that a site has
$k$ neighbours decays as $1/k^3$ for $k \ge m$.

If one now wants to put opinion dynamics onto this network, one may waste 
much memory. With a million sites it is possible that one of them (typically
one of the starting sites) has a large number of neighbours, of the order 
thousands. Then a neighbourhood table of size $10^6 \times 10^4$ is needed.
Aleksiejuk \cite{aleks} programmed a one-dimensional neighbourhood table to save
memory, but this is difficult to understand. It is much more practical, and 
does not change the results much, to switch from the above undirected networks 
to directed networks: The new site still selects $m$ neighbours from the old
sites, but these $m$ old sites do not have the new site as a neighbour: The
neighbourhood relation has become asymmetric or directed. Similarly, a new
member of a political party knows the heads of that party, but these heads 
don't know the new member. Thus a hierarchy of directed relations is built 
up, which starts with the latest members of the network at the bottom and
ends with the initial core members at the top.

Actually, Figs.6,7 are for a directed scale-free network, but the undirected 
case with infinite-range connectivity looks very similar \cite{fort3} for
opportunists. The simplicity of the scaling law makes it invariant against 
details of the network.  

Also for other questions \cite{hmo}, negotiators with continuous opinions
between zero and one on a Barab\'asi-Albert network showed little difference
between directed and undirected neighbourhoods. Fig.8 shows the resulting 
size distribution of opinion clusters. Such an opinion cluster is the set of 
people, not necessarily connected, having the same (within $10^{-6}$)
opinion. Consensus means everybody is in one opinion cluster; thus the
opposite fragmentation limit $\epsilon=0.1$ is more interesting. If we increase
systematically the network size, we see in Fig.8 a peak moving to the right 
with increasing network size. This peak comes from large clusters, of the order
of one per simulated network, which contain a positive fraction of the whole
network. To the left of this peak, separated by a valley only for large 
networks, is the statistics for lots of small clusters, down to isolated 
people sharing their opinion with nobody else.  The size distribution
of these many small
clusters is not much affected by the network size except that their statistics 
is better for larger networks. This cluster statistics is similar to percolation
theory slightly above the threshold: One ``infinite'' cluster coexists with
many finite clusters. But our clusters are sets of agreeing people, 
not sets of neighbouring sites as in percolation theory.

\begin{figure}[hbt]
\begin{center}
\includegraphics[angle=-90,scale=0.5]{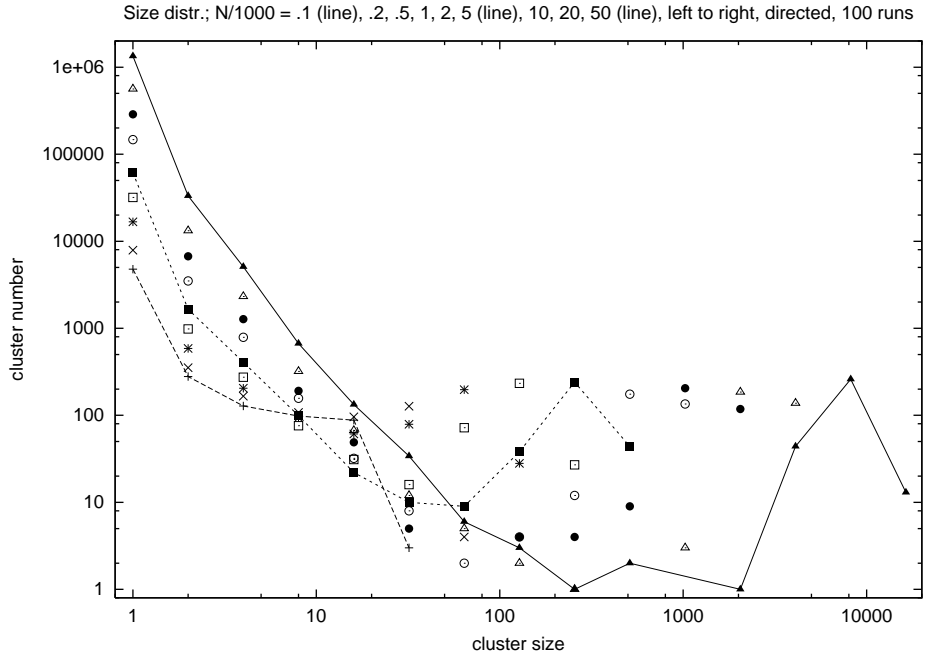}
\end{center}
\caption{Size distribution of opinion clusters for negotiators on directed
Barab\'asi-Albert networks, with continuous opinions between zero and one
and $\epsilon=0.1$. All network sizes, from 100 to 50,000, were simulated
100 times. From \cite{deffdis}.
}
\end{figure}

\begin{figure}[hbt]
\begin{center}
\includegraphics[angle=-90,scale=0.5]{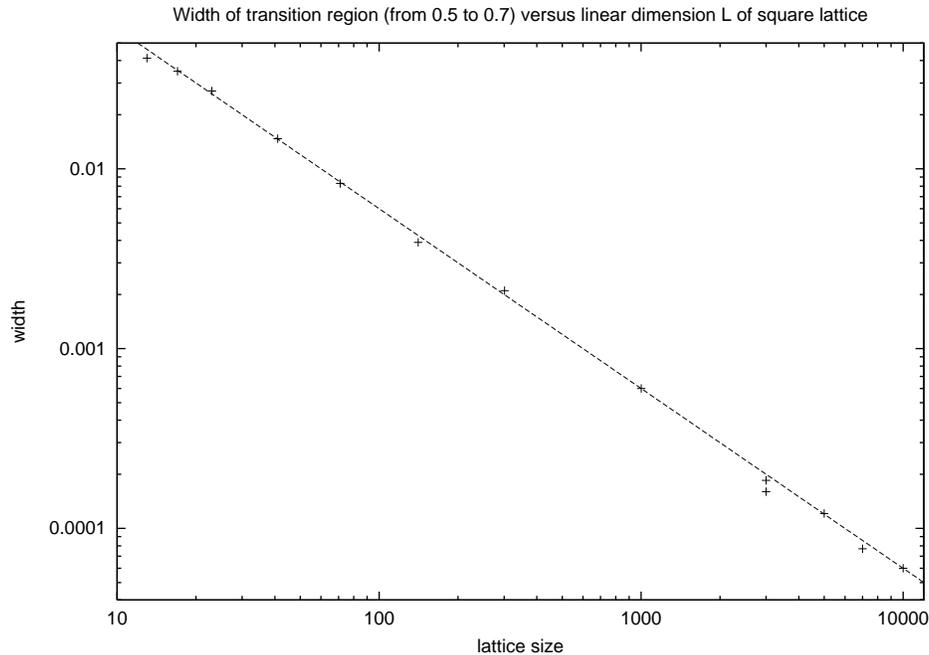}
\end{center}
\caption{Transition width for missionaries with mixed global and
local interaction, from \cite{schulzegl}. This width is defined as the 
interval in the initial concentration of up opinions ($Q=2$ square lattice)
within which the fraction of samples finishing all up increases from 50 to 70 
percent.
}
\end{figure}

\section {Phase Transition} 
The simplest version of missionaries \cite{sznajd} has only $Q=2$ possible 
opinions up and down, in which case bounded confidence makes no sense and 
can be omitted. We always reach a consensus on the square lattice: Either 
everybody is up or everybody is down. If initially half of the opinions are
up and the others are down, then after the simulation of many samples one
finds \cite{moss} that half of the samples are all up and half are all down. 
(For negotiators, the limit $Q=2$ destroys the compromise part of shifting 
opinions somewhat and thus does not make sense. For opportunists we expect 
in this limit the same behaviour as for missionaries except that the dynamics
is much faster: Each person takes the initial majority opinion.) 

When we vary the initial up concentration away from 50 percent, then for the
small and intermediate sizes for which the standard missionary model reaches
its final fixed point within reasonable time, the fraction of all-up samples
also moves away from 50 percent. Increasing the initial up concentration from 
40 to 60 percent, we see \cite{moss} for small square lattices a slow increase
of the fraction of up-samples, and for larger lattices this increase is 
somewhat steeper. Thus one might extrapolate that for infinite lattices 
one has a sharp phase transition: For initial up concentrations below 50 percent
all samples finish down, and for concentrations above 50 percent all samples 
finish up. For the standard model the numerical evidence is meager due to small
system sizes, but Schulze \cite{schulzegl} combined the traditional 
nearest-neighbour interactions with global interactions similar to the nice
mean-field theory of of Slanina and Lavi{\c c}ka \cite{slanina}: 
two people of arbitrary distance who agree in their
opinions convince their nearest neighbours of this opinion. Then, as predicted
\cite{slanina}, the times to reach the final fixed point are much shorter and 
their distribution decays exponentially. Thus larger lattices can be simulated
and give in Fig.8 a width of the transition varying as $1/L$ in $L \times L$
squares, $10 < L < 10,000$. This is clear evidence for a simple phase 
transition.

In the simplification \cite{ochrombel} suggested by a third-year student,
to allow already one single missionary to convince the four square-lattice
neighbours, one still has a complete consensus but no more a phase transition:
The fraction of final up samples agrees with the initial concentration of
up opinions, as found independently in simulation \cite{schulzead} and 
approximate theory \cite{slanina}; see Slanina lecture in this Seminar.
Easier to understand is that in one dimension, also this phase transition
does not exist \cite{sznajd}; this absence corresponds to the lack of phase 
transition in the one-dimensional Ising model of 1925 which is so often 
mispronounced as Eyesing-Model by unhistorical speakers, instead of Eesing.
 
\section{Variants}

Numerous variants of the above standard models were published, and can be 
summarized here only shortly.

{\bf Negotiators:} 
The inventors of negotiators \cite{deffuant} published several alternative, 
for example with unsymmetric opinion shifts as a result of compromise, 
with negotiators on a square lattice etc \cite{deffuant2}. For negotiators on
scale-free networks, the network was made more realistic by increasing 
triangular relations (the friend of my neighbours is also my own friend);
the qualitative results remained unchanged \cite{deffdis}. Opinions which differ
in their convincing power were simulated by Assmann \cite{assmann}, and 
the interactions of opinions on several different aspects of life by 
Jacobmeier \cite{jacob}.
 
{\bf Missionaries:}
If you dislike the complete consensus enforced by missionaries, just let their
neighbours follow the convincing rule only with a high probability, or let a 
small minority of dissidents always be against the majority. Then the full
consensus is replaced by the more realistic widespread consensus 
\cite{sznajd,schneider}. The role of neutrality in a three-opinion model,
of opinion leaders, and of social temperature was
studied by He et al \cite{he}. Sousa \cite{sousa} showed that the network
results are robust against the inclusion of more triangular relations
(preceding paragraph) and that complete consensus can be avoided with $Q > 2$.
If the opinion dynamics starts already while the network is still growing
not much is changed\cite{bonnekoh}.

According to \cite{behera}, the missionaries are part of a wider group of
cellular automata giving about the same results; see also \cite{galam}. 
Long-range interactions, decaying with a power law of the distance on a 
square lattice, still need to be explained \cite{schulzelr}.
Frustration occurs if we switch from sequential to simultaneous updating 
and one person gets different opinions simultaneously from different pairs 
of missionaries \cite{frustration}.
The time-dependent decay of the number of people who never changed their opinion
is Ising-like only in one dimension \cite{pmco}. Other (dis-)similarities with
Ising models are discussed in \cite{sznajdising,brazil}.

{\bf Opportunists:} Fortunato \cite{fort2} compared the threshold for $\epsilon$
when the number of neighbour varies proportional to the total population to the
case where it is independent of the population size. Hegselmann and Krause
\cite{hegs} compared various alternative averages to the standard 
arithmetic average.

\section{Applications}

The most successful application of the missionary model were political 
elections. This does not mean that we can predict which candidate will
win the next elections. Neither can statistical physics predict which air 
molecules will hit my nose one minute from now; the laws of Boyle-Mariott and 
Gay-Lussac predict the pressure, i.e. the average number of molecules hitting
my nose per picosecond. Similarly, many elections have shown a rather similar 
picture for  the number of votes which one candidate gets (in case voters can
select among numerous candidates, not among a few parties). The larger the
number $v$ of votes is, the smaller is the number $n(v)$ of candidates getting
$v$ votes. For intermediate $v$ one has $n(v) \propto 1/v$ while for large and 
small $v$, downward deviations are seen: Nobody gets more than 100 percent of 
the 
votes, and nobody gets half a vote. Missionaries on their way to the consensus 
fixed point on a Barab\'asi-Albert network agreed well with Brazilian votes
\cite{kertesz}, and similar agreement was also found in modified networks
\cite{gonzalez,sousa} and Indian elections. However, exceptions exist (S. Fortunato, priv. comm.). It
would be interesting to check whether opportunists and negotiators also agree
with Brazilian election results.

If one person changes opinion, does this influence the whole community ? This
question, known as damage spreading in physics but invented in 1969 for 
genetics by Stuart Kauffman, was recently simulated in detail by Fortunato
and the review \cite{forsta} is still up-to-date.

Readers may try to become rich by going into advertising: How can mass media
influence opinion dynamics? For missionaries \cite{schulzead} the answer is
clear: The larger the population is the less effort is needed to convince
everybody to drink Coke instead of Pepsi; but the advertizing has to come early
in the opinion formation process, not when most people have already made
their choice. Again, analogous studies for opportunists and negotiators would
be nice. Or perhaps you get rich with \cite{weron}.

\section{Summary}

Humans may dislike to be simulated like Ising spins, and clearly the brain is
more complicated than one binary variable. But humans have been treated in
this way since a long time: The astronomer Halley, known for his comet, tried
to estimate human mortality already three centuries ago. Life insurance,
health insurance, car insurance are present widespread examples of treating
humans like inanimate particles with probabilistic behaviour, relying on the
laws of large numbers. Whoever dislikes this treatment, should not blame todays
sociophysicists for having started it. Already more than two millenia ago,
Empedocles compared humans with fluids: Some are like
wine and water, mixing well; others dislike each other, like oil and water
(J. Mimkes, priv. comm.).

\section{Summary} 
 
%\begin{theacknowledgments}
I thank my collaborators working on these models since the beginning of
this millenium: S. Moss de Oliveira, A.O. Sousa, J.S. Andrade, A.A. Moreira,
A.T. Bernardes, U.M.S. Costa, A. Araujo, R. Ochrombel, C. Schulze, J. Bonnekoh, 
P.M.C de Oliveira, H. Meyer-Ortmanns, S. Fortunato, P. Assmann, and N. 
Klietsch.
%\end{theacknowledgments}

\end{document}